\documentclass[12pt,showpacs,aps,prb,preprint]{revtex4}   % style for Physical Review B and AJP are similar

\usepackage{amsmath}    % need for subequations
\usepackage{graphicx}   % for figures
\graphicspath{{../figures/}}

\begin{document}

\title{The meaning of 1 in {\it j(j+1)}

El significado del 1 en {\it j(j+1)}}
\author{E. Gomez}
 \affiliation{Instituto de F{\'{i}}sica, Universidad Aut{\'{o}}noma de San Luis Potosi,
San Luis Potosi 78290}
 \email{egomez@ifisica.uaslp.mx}   %optional
\date{\today}

\begin{abstract}
The magnitude of the angular momentum ($J^2$) in quantum mechanics is larger than expected from a classical model. We explain this deviation in terms of quantum fluctuations. A standard quantum mechanical calculation gives the correct interpretation of the components of the angular momentum in the vector model in terms of projections and fluctuations. We show that the addition of angular momentum in quantum mechanics gives results consistent with the classical intuition in this vector model.

La magnitud del momento angular ($J^2$) en mec\'{a}nica cu\'{a}ntica es mas grande que lo esperado en un modelo cl\'{a}sico. Explicamos esta diferencia en t\'{e}rminos de las fluctuaciones cu\'{a}nticas. Un c\'{a}lculo est\'{a}ndar de mec\'{a}nica cu\'{a}ntica da la interpretaci\'{o}n correcta a las componentes del momento angular en el modelo vectorial en t\'{e}rminos de proyecciones y fluctuaciones. Mostramos que la suma de momento angular en mec\'{a}nica cu\'{a}ntica da resultados consistentes con la intuici\'{o}n cl\'{a}sica en este modelo vectorial.
\end{abstract}

\maketitle

\section{Introduction}

The operator of angular momentum in quantum mechanics is always a confusing topic for new students. The quantum description of angular momentum involves differential operators or new algebra rules that seem to be disconnected from the classical intuition. 
For small values of angular momentum one needs a quantum description because the quantum fluctuations are as big as the angular momentum itself. In this regime, the simple classical models generally do not give the right result. In this paper I describe the use of fluctuations in the angular momentum components to produce a vector model compatible with the quantum mechanical result. I show that the addition of angular momenta from a standard quantum mechanical calculation is consistent with the classical intuition using the vector model. The paper is organized as follows: Section \ref{vector} shows the problems encountered with the vector model, section \ref{spin} works out the details for a spin 1/2 particle, section \ref{addition} explains the addition of angular momenta for two spin 1/2 particles, section \ref{general} describes the general case of addition of angular momenta and I give some conclusions at the end.

\section{\label{vector}Vector model of angular momentum}

The presentation of angular momentum in quantum mechanics textbooks demonstrates the following relations \cite{sakurai}
\begin{subequations} 
\begin{align}
\langle J^2 \rangle &= j(j+1)\hbar^2
\label{magnitudej} \\
\langle J_z \rangle &= m\hbar,
\label{projectionj}
\end{align}
\end{subequations}
with $-j \leq m \leq j$. It is usually said that the angular momentum comes in units of $\hbar$. This is consistent with Eq.~\ref{projectionj} since $m$ is an integer, but not with Eq.~\ref{magnitudej}. For example, if we have one unit of angular momentum ($j=1$), then $\langle J^2 \rangle = 2\hbar^2$, that is, the magnitude of the angular momentum is $\sqrt{2}$ rather than 1 (from now on we express angular momentum in units of $\hbar$). Only the $z$ projection of the angular momentum comes in units of $\hbar$ and not the magnitude. How can we reconcile both expressions? There is a nice derivation that explains the expression for $J^2$ by averaging the value of $J_z^2$.\cite{mcgervey91,milonni89,feynman} There are also ways to give an heuristic derivation of the properties of angular momentum.\cite{leblond76} We would like to gain some intuition as to where the extra 1 in Eq.~\ref{magnitudej} comes from.

The vector model is often introduced to give a classical analogy to the quantum angular momentum.\cite{cohentannoudji} To describe the angular momentum classically by a vector, we must specify its three components ${\cal J}_x$, ${\cal J}_y$ and ${\cal J}_z$. The magnitude of the vector is obtained from the components. The problem with that scheme in quantum mechanics is that it is impossible to measure with absolute precision the three components of the angular momentum. If one measures $J_z$ and $J_y$ exactly then the uncertainty in $J_x$ grows, that is, there is an uncertainty relation for the components of the angular momentum analogous to the uncertainty relation between position and momentum.

The natural choice for the components of angular momentum in the vector model would be ${\cal J}=(\langle J_x \rangle,\langle J_y \rangle,\langle J_z \rangle)$. We will show that this choice (choice A) gives the incorrect value for ${\cal J}^2$. A better choice (choice B) for the angular momentum vector is ${\cal J}=(\langle J_x^2 \rangle^{\frac{1}{2}},\langle J_y^2 \rangle^{\frac{1}{2}},\langle J_z^2 \rangle^{\frac{1}{2}})$. With this choice the magnitude square of the angular momentum vector gives the correct value ${\cal J}^2=\langle J_x^2 \rangle + \langle J_y^2 \rangle + \langle J_z^2 \rangle$. In the next section we give a classical interpretation of the components of the angular momentum vector in terms of fluctiations and we use this interpretation to explain the origin of the extra 1 in Eq.~\ref{magnitudej}.

\section{\label{spin}Spin 1/2 case}

The key point to explain Eq.~\ref{magnitudej} lies in the fluctuations. Take the case of a state with spin 1/2 ($j=s=1/2$) and $m_s=1/2$. The values of $\langle S_x \rangle$, $\langle S_y \rangle$ and $\langle S_z \rangle$ are 0, 0 and 1/2 respectively. Choice A for the vector model gives ${\cal S}=(0,0,1/2)$ and the magnitude square of this vector is ${\cal S}^2=1/4$, which differs from the result $\langle S^2 \rangle=3/4$ obtained from Eq.~\ref{magnitudej}. Choice B gives right value for ${\cal S}^2$ since it was constructed that way. What is the meaning of each component? ${\cal S}_z=\langle S_z^2 \rangle^{\frac{1}{2}}=(\langle S_z \rangle^2)^{\frac{1}{2}}=\langle S_z \rangle$ and this component reduces to the $z$ projection of the operator S. For ${\cal S}_x$ we cannot use the same trick since we are not using an eigenstate of $S_x$. Still we can relate that component to the fluctuations. The fluctuations of an operator $A$ in quantum mechanics are given by \cite{sakurai}
\begin{equation}
\Delta A^2 = \langle A^2 \rangle - \langle A \rangle^2.
\label{variance}
\end{equation}
For the present state and the operator $S_x$ the result is
\begin{equation}
\Delta S_x^2 = \langle S_x^2 \rangle.
\label{uncertaintys}
\end{equation}
Then ${\cal S}_x=\langle S_x^2 \rangle^{\frac{1}{2}}=(\Delta S_x^2)^{\frac{1}{2}}=\Delta S_x$ and this component is equal to the fluctuations in the $x$ axis of the operator $S$. The $y$ component gives the same result. The meaning of the vector components in choice B is that ${\cal S}_x$ and ${\cal S}_y$ are fluctuations and ${\cal S}_z$ is the projection in the corresponding axis. The quantum mechanical calculation of the fluctuations gives
\begin{equation}
\Delta S_x^2=\langle \frac{1}{2} | ~S_x^2~ | \frac{1}{2} \rangle = \frac{1}{4}, \label{fluctuations12}
\end{equation}
then $\Delta S_x=1/2$, and similarly $\Delta S_y=1/2$. The vector is ${\cal S}=(1/2,1/2,1/2)$ and the magnitude square of the vector is ${\cal S}^2=3/4$ which is the correct value. The value of ${\cal S}^2$ in choice A is $s^2=1/4$. Instead in choice B, ${\cal S}_x$ and ${\cal S}_y$ contribute to ${\cal S}^2$ through the fluctuations giving the value of $s(s+1)=3/4$.

\section{\label{addition}Addition of angular momenta}

We construct any value of angular momentum by adding several spin 1/2 particles. We show how the vector model works for two spin 1/2 particles. The sum of two spin 1/2 particles gives a total angular momentum of $j=1$ or $j=0$. Take first the case of the state with $j=1$ and $m=1$. The state is represented in quantum mechanics by $|\frac{1}{2},\frac{1}{2}\rangle$ where the numbers represent the $z$ projection of the spin of particles 1 and 2 respectively. The objective is to calculate the value of $\langle J^2 \rangle$, with $J=S_1+S_2$, the sum of the spin contributions. The quantum mechanical result from Eq.~\ref{magnitudej} is $\langle J^2 \rangle=2$, and we want to explain this in terms of the vector model.

The expression for $J^2$ is
\begin{equation}
J^2 = J_x^2 + J_y^2 + J_z^2 = (S_{x1}+S_{x2})^2 + (S_{y1}+S_{y2})^2 + (S_{z1}+S_{z2})^2,
\label{jdecomposed}
\end{equation}
where the index 1 and 2 refer to particles 1 and 2 respectively. There is no question as to how to calculate the expectation values in quantum mechanics, but if we think in terms of the vector model we are in trouble since we have to add two vectors that are a mixture of projections and fluctuations. We show the correct recipe for adding this vectors from a quantum mechanical calculation and show that it is consistent with the classical intuition.

Take ${\cal J}_z$ first. The sum is again simplified since we use an eigenstate of the operator. We have ${\cal J}_z=(\langle (S_{z1}+S_{z2})^2 \rangle)^{\frac{1}{2}} = \langle S_{z1} \rangle + \langle S_{z2} \rangle$, that is, ${\cal J}_z$ is just the direct sum of the individual projections. The $x$ component gives
\begin{equation}
{\cal J}_x=(\langle (S_{x1}+S_{x2})^2 \rangle)^{\frac{1}{2}} = (\langle S_{x1}^2 \rangle + 2 \langle S_{x1}S_{x2} \rangle + \langle S_{x2}^2 \rangle)^{\frac{1}{2}} = (\Delta S_{x1}^2 + \Delta S_{x2}^2)^{\frac{1}{2}}.
\label{xsumspin}
\end{equation}
The two contributions add up in quadrature. This is expected since the $x$ component for each spin in the vector model corresponds to fluctuations (or noise), and the proper way to add uncorrelated noise is in quadrature. For a classical variable $w=u+v$, where $u$ and $v$ are fluctuating variables, the noise in $w$ is given by \cite{bevington}
\begin{equation}
\sigma_w^2 = \sigma_u^2 + \sigma_v^2 + 2 \sigma_{uv}^2.
\label{noisefullformula}
\end{equation}
The quantum mechanical expression for the fluctuations of $J_x=S_{x1}+S_{x2}$ for the present state is
\begin{equation}
\Delta J_x^2 = \langle S_{x1}^2 \rangle + \langle S_{x2}^2 \rangle + 2\langle S_{x1}S_{x2} \rangle,
\label{noisesxsum}
\end{equation}
where the similarity between the last two expressions is evident. The state we are considering has the two spins aligned. Since the two spins are independent, we expect their noise to be uncorrelated. The calculation of the correlation term (last term in Eq.~\ref{noisesxsum}) gives
\begin{equation}
\langle \frac{1}{2},\frac{1}{2}| ~S_{x1}S_{x2}~ |\frac{1}{2},\frac{1}{2}\rangle = 0, \label{correlationterm}
\end{equation}
and the sum for ${\cal J}_x$ reduces to Eq.~\ref{xsumspin}.

We can understand the addition of angular momentum in the vector model: the components that are projections add up directly whereas the components that are fluctuations add up as noise. The noise can have different degrees of correlation as calculated by the last term in Eq.~\ref{noisesxsum}. The noise for the present state happens to be uncorrelated (Eq.~\ref{correlationterm}). The vectors for the individual spins are ${\cal S}_1={\cal S}_2=(1/2,1/2,1/2)$ and their sum gives ${\cal J}=(1/\sqrt{2},1/\sqrt{2},1)$ where we have added the $x$ and $y$ components in quadrature and the $z$ components directly. The magnitude square of the vector gives ${\cal J}^2=2$ in accordance with Eq.~\ref{magnitudej}. The result should be contrasted with a naive addition of the vectors ${\cal S}_1+{\cal S}_2=(1,1,1)$, that gives a magnitude square of 3. The case for the state with $j=1$ and $m=-1$ works the same way. 

The state with $j=1$ and $m=0$ is more interesting. The state is the symmetric combination of the spins, $(|\frac{1}{2},-\frac{1}{2} \rangle + |-\frac{1}{2},\frac{1}{2} \rangle)/\sqrt{2}$. The vectors for the individual spins are ${\cal S}_1=(1/2,1/2,1/2)$ and ${\cal S}_2=(1/2,1/2,-1/2)$. We take the negative value of the square root in ${\cal S}_{2z}$ since the $z$ component of the two spins point in opposite directions. We choose ${\cal S}_{1z}$ (${\cal S}_{2z}$) positive (negative), but the opposite is equally correct. In the direct sum of the $z$ components ${\cal S}_{1z}$ and ${\cal S}_{2z}$ cancel each other giving 0. The correlation term in the $x$ component for this state gives
\begin{equation}
\frac{1}{\sqrt{2}} \left( \langle \frac{1}{2},-\frac{1}{2}|+ \langle -\frac{1}{2},\frac{1}{2}| \right) ~S_{x1}S_{x2}~ \left( |\frac{1}{2},-\frac{1}{2}\rangle + |-\frac{1}{2},\frac{1}{2}\rangle \right) \frac{1}{\sqrt{2}} = \frac{1}{4}, \label{correlationterm2}
\end{equation}
and the calculation for the fluctuations from Eq.~\ref{noisesxsum} gives $\Delta J_x^2=\Delta J_y^2=1$. The sum vector is ${\cal J}=(1,1,0)$ that gives the correct result for the magnitude square ${\cal J}^2=2$. The noise calculation for ${\cal J}_x$ and ${\cal J}_y$ tells us that we have perfectly correlated noise, so instead of adding the two contributions in quadrature, we add them directly (each one equal to 1/2 giving a total of 1). It is not unexpected that the noise behaves in a correlated manner since we use the symmetric combination of the spins.

Finally the case with $j=0$ and $m=0$. The state is the anti-symmetric combination of the spins and we expect the noise to be anti-correlated. The vectors for the indivicual spins are still ${\cal S}_1=(1/2,1/2,1/2)$ and ${\cal S}_2=(1/2,1/2,-1/2)$. The correlation term in the $x$ component for this state is now
\begin{equation}
\frac{1}{\sqrt{2}} \left( \langle \frac{1}{2},-\frac{1}{2}|- \langle -\frac{1}{2},\frac{1}{2}| \right) ~S_{x1}S_{x2}~ \left( |\frac{1}{2},-\frac{1}{2}\rangle - |-\frac{1}{2},\frac{1}{2}\rangle \right) \frac{1}{\sqrt{2}} = -\frac{1}{4}, \label{correlationterm3}
\end{equation}
and the calculation for the fluctuations from Eq.~\ref{noisesxsum} gives $\Delta J_x^2=\Delta J_y^2=0$. The sum vector is ${\cal J}=(0,0,0)$ with a magnitude square of ${\cal J}^2=0$ as expected. The anti-symmetric combination of the spins results in noise that is perfectly anti-correlated (due to the minus sign in the wave function). The noise subtracts directly ($\frac{1}{2}-\frac{1}{2}=0$) and not in quadrature for the $x$ and $y$ components. It seems that the noise in $J_x$, $J_y$ and $J_z$ is zero for the state. From the point of view of the sum, the individual perpendicular fluctuations are actually not zero, it is because of the correlations that the fluctuations of $J$ become zero.

\section{\label{general}General case}

Any other value of angular momentum can be constructed using the same scheme. For example, to obtain $j=3/2$ we add three spin 1/2 particles. Each particle contributes some amount to the value of ${\cal J}_z$ and also to the fluctuations in the perpendicular components. There is some degree of correlation between the spins depending on the $m$ value chosen. The correlation between spins can be calculated from the crossed term in Eq.~\ref{noisesxsum}. The correlation term between spins $i$ and $k$ in the x component for the state with angular momentum $j$ and projection $m$ is
\begin{equation}
\langle j,m| ~S_{xi}S_{xk}~ |j,m \rangle. \label{correlationtermij}
\end{equation}
It is not trivial to predict the result of this calculation except for the maximum and minimum projections. All the spins are uncorrelated if $m=j$ or $m=-j$. For any other projections there will be some intermediate degree of correlation between spins that can be calculated from Eq.~\ref{correlationtermij}. For the maximum projection, the $x$ (and $y$) components of all the individual spins add up in quadrature to give
\begin{equation}
{\cal J}_x=\sqrt{\Delta S_{x1}^2+\Delta S_{x2}^2+...+\Delta S_{x(2j)}^2}=\sqrt{2j(1/4)}=\sqrt{j/2}. \label{generalsum}
\end{equation}
The vector sum is ${\cal J}=(\sqrt{j/2},\sqrt{j/2},j)$ with a magnitude square ${\cal J}^2=j/2+j/2+j^2=j(j+1)$, where the 1 comes from the perpendicular components.

\section{\label{conclusions}Conclusions}

We explain the 1 in the expression $j(j+1)$ in terms of the quantum fluctuations of the $x$ and $y$ components of the angular momentum. We include the fluctuations to describe the addition of angular momenta in the vector model. The vector components can be projections or fluctuations and they have different formulas for addition. The correlations in the fluctuations cannot be ignored. Formula \ref{magnitudej} tells us that angular momentum does not come in units of $\hbar$, but instead it comes in units of $\sqrt{1+(1/j)}\hbar$. This is not even a uniform unit, but depends on the value of $j$ in a complicated way. This happens because some of the components of $J$ add up directly and others in quadrature. Only in the limit of big $j$ we recover the well known $\hbar$ unit of angular momentum. At small $j$ the quantum noise cannot be ignored.

\begin{acknowledgments}
I would like to thank A. P{\'{e}}rez, E. Ugalde and J. Ur{\'{i}}as for helpful discussions. 
\end{acknowledgments}


\begin{thebibliography}{8}

\bibitem{sakurai}J. Sakurai, {\it Modern Quantum Mechanics} (Addison Wesley, Massachusetts, 1994), pp. 34-36, 187-191, 217-221.

\bibitem{mcgervey91}J. McGervey, {\it Am. J. Phys.} {\bf 59} (1991) 295-296.

\bibitem{milonni89}P. Milonni, {\it Am. J. Phys.} {\bf 58} (1989) 1012.

\bibitem{feynman}R. Feynman, R. Leighton and M. Sands, {\it The Feynman Lectures on Physics}
(Addison Wesley, Massachusetts, 1965), Vol. II, pp. 34-11.

\bibitem{leblond76}J. Levy-Leblond, {\it Am. J. Phys.} {\bf 44} (1976) 719--722.

\bibitem{cohentannoudji}C. Cohen-Tannoudji, B. Diu and F. Laloe, {\it Quantum Mechanics}
(Wiley-Interscience, New York, 1977), Vol I, pp. 668-670. 

\bibitem{bevington}P. Bevington and D. Robinson, {\it Data reduction and error analysis for the physical
sciences} (McGraw-Hill, Massachusetts, 1992), pp. 38-52.

\end{thebibliography}
\end{document}